\documentclass{menu99}    
\usepackage{epsfig}

\begin{document}
    \setlength{\baselineskip}{2.6ex}
\begin{flushright}
HIP -- 1999 -- 75/TH
\end{flushright}
\title{Working Group Summary: Pion-Nucleon Coupling Constant}
\author{M.E. Sainio \\
{\em Helsinki Institute of Physics {\rm and}\\
Department of Physics, University of Helsinki\\
P.O.B. 9, FIN-00014 Helsinki, Finland}}

\maketitle

\begin{abstract}
\setlength{\baselineskip}{2.6ex}
A brief introduction to different determinations
of the $\pi$NN coupling constant is given, and some comments on the
topics discussed in the working group are made.
\end{abstract}

\setlength{\baselineskip}{2.6ex}

\section*{INTRODUCTION}

Since the birth of the Yukawa theory of the nuclear force in 1935 it was 
a challenge for the physics community to determine the coupling strength
of the Yukawa meson to the nucleon. In 1947 the meson -- pion -- was finally discovered in 
cosmic ray emulsion experiments \cite{1} and more systematic work to determine
the $\pi$NN coupling constant could start. Conventionally \cite{2} the pseudoscalar 
strength is denoted by $g$ and the pseudovector coupling constant by $f$ such that
\begin{eqnarray}
f^2 = \left(\frac{M_\pi}{2 m_p}\right)^2 \; \frac{g^2}{4 \pi},
\end{eqnarray}
where $M_\pi$ is the charged pion mass and $m_p$ is the proton mass.
Other conventions concerning the nucleon mass and the factor $4 \pi$ appear in the
literature \cite{3}.
Reasonable estimates for the coupling strength were obtained even before the 
discovery of the pion and 
without detailed knowledge of the meson mass, e.g., Bethe was able
to get an estimate $f^2$ = 0.077 - 0.080 already in 1940 \cite{4} on the basis 
of deuteron properties. The results of various determinations
until 1980 are shown in Fig. 1. In the same figure very different
techniques to determine $f^2$ are summarized. In the previous {\em MENU}
symposium de Swart gave a review on the topic \cite{3} and many
of the references used in Fig. 1 can be found there.
\begin{figure}[h]
\begin{center}
\mbox{\epsfysize=8cm \epsfbox{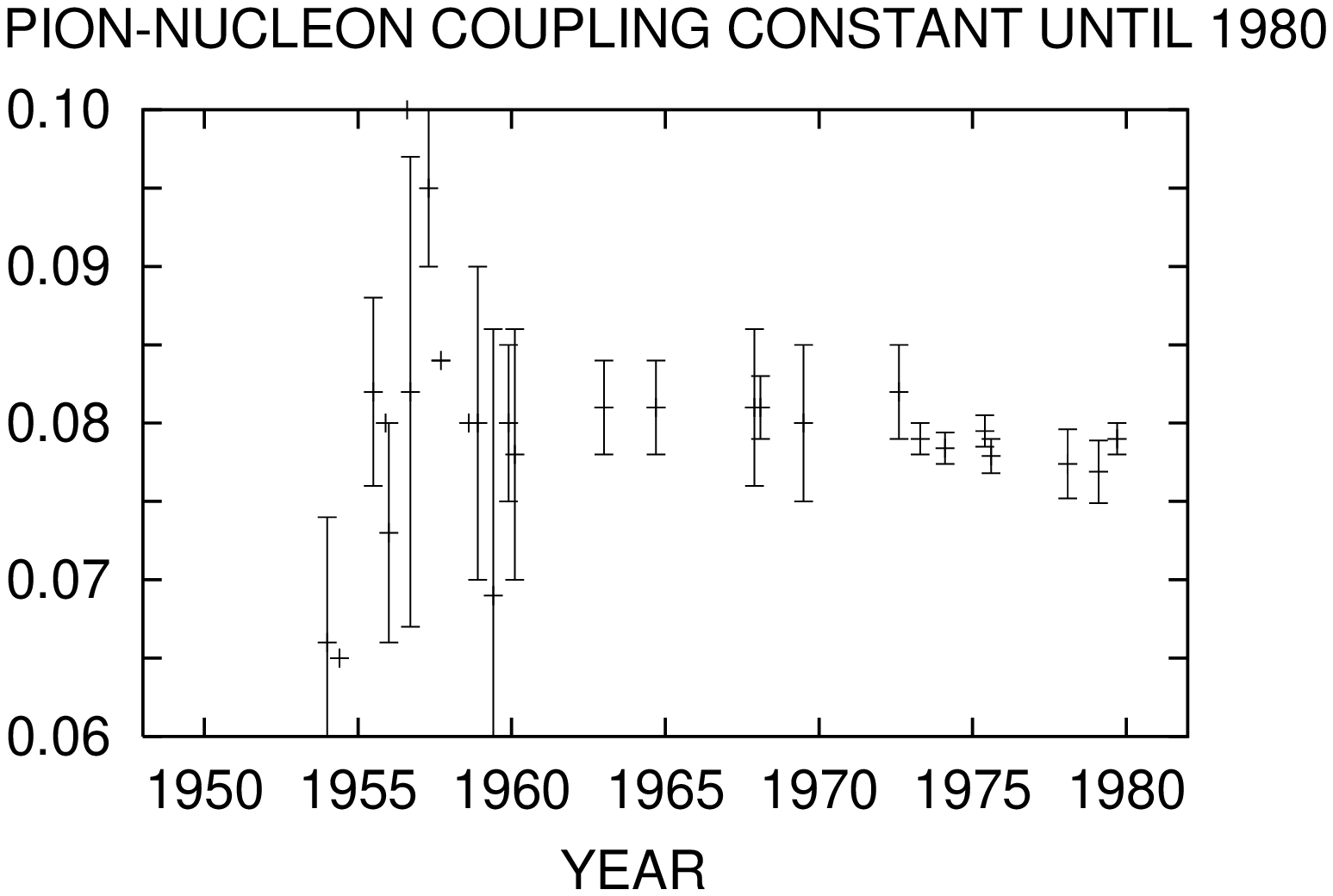}
     }
\caption{The values of the pion-nucleon coupling constant $f^2$ before 1980.}
\label{fig1} \end{center} \end{figure}
The values of $f^2$ stabilized for a long time \cite{5,6,7} and only in the 90's
has the discussion of the value of the pion-nucleon coupling constant
started again. In \cite{5,6,7} fixed-$t$ dispersion relations 
for $\pi$N were used. In the determinations displayed in Fig. 1 most of the data
date back to the era before the meson factories, LAMPF, SIN and TRIUMF, which,
in addition to performing experiments with pions, had programmes to study the
NN interaction.
\begin{figure}[t]
\begin{center}
\mbox{\epsfysize=8cm \epsfbox{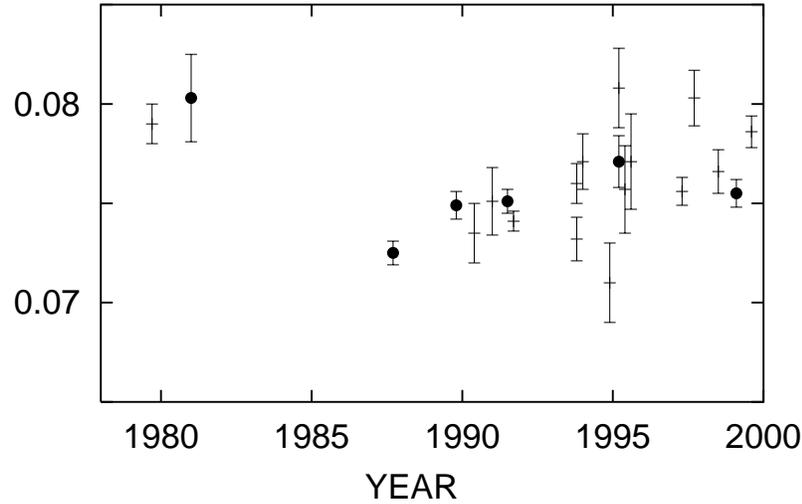}
     }
\caption{The values of the pion-nucleon coupling constant $f^2$ after 1980
until the present. Neutral pion couplings are denoted by the solid dots, the
remaining points refer to charged pion couplings or charge independent
determinations. }
\label{fig2} \end{center} \end{figure}
In several analyses shown in Fig. 2 NN scattering
data have been used to extract the $\pi$N coupling strength, i.e. one of the
standard methods until the 60's has been adopted again in more refined form.
In this activity the Nijmegen group has played an important role \cite{3}.
Of course, in Fig. 2  many results from the meson factory $\pi$N experiments are
included as well in the data bases used to determine $f^2$.

The central issue in the discussion in the working group has been the scatter
of the results of the determinations as shown in Fig. 2. The main questions
involve the model dependence of different techniques, the effect of different
pieces of data (partly conflicting), the error estimates, the electromagnetic 
corrections and other isospin violating effects. In the working group 
contributions were presented by Loiseau \cite{8}, H\"ohler \cite{9} and 
Pavan \cite{10}, and brief commentaries
by W.R. Gibbs, M. Birse and D.V. Bugg.

\section*{PROBLEMS IN EXTRACTING $f^2$}

The particular issues raised in the discussion include:
\begin{itemize}
\item Electromagnetic corrections:
\begin{itemize}
\item $\pi$N vs. NN; the different treatment of electromagnetic
corrections for these two scattering processes gives a possibility
to check the uncertainty due to these effects
\item corrections from Tromborg et al. \cite{11} vs. Oades et al. \cite{12};
the dispersion approach and potential model lead to differences which need
checking
\item corrections at high energy; these need to be checked, Tromborg et al.
calculated corrections only up to 655 MeV/c 
\end{itemize}
\item Lack of transparency of the analyses; the analyses contain large
data bases and it is hard to clarify which pieces of information are
the crucial ones in determining $f^2$
\item The normalization of the $np$ data is a problem in the $(p,n)$ data
analyses
\item The determination of the s-wave isoscalar scattering length, $a^+_{0+}$,
from the $\pi^- d$ level shift measurement suffers from  some model dependence
due to electromagnetic and absorption corrections
\item Effective theory is not at present suitable for fixing the coupling constant.
The problems relate mostly to the convergence of the chiral expansion or to
the additional low-energy constants which are not known accurately enough.
However, there might be a chance in a precise measurement of the induced pseudoscalar 
coupling constant, \newline
$g_P$, which would make an accurate determination of the pion-nucleon
coupling constant possible \cite{13}
\item There is need for a new fixed-$t$ analysis of $\pi$N scattering data which
extends beyond the present limit of the VPI analysis, 2.1 GeV
\item There is need for a new analysis of the forward dispersion relations of the
NN system. The amount of NN data has increased considerably since the previous
analysis thanks to the meson factories and SATURNE.
\end{itemize}

\subsection*{The GMO Sum Rule}

The Goldberger-Miyazawa-Oehme sum rule (GMO) \cite{14} provides a simple means
to estimate the pion-nucleon coupling constant directly from measurable
quantities, the  $\pi$N isovector s-wave scattering length and total
cross sections from the threshold to the highest energies. The method
still has uncertainties, and will probably never be able to compete
with other methods in precision, but the advantage is the possibility
to relate the uncertainty in $f^2$ directly to the experimental errors.

The GMO sum rule is the result of the forward dispersion relation for
the \newline
$D^- (=A^- + \nu B^-)$ amplitude taken at the physical threshold 
(the total laboratory energy $\omega = M_\pi$)
\begin{eqnarray}
D^-(M_\pi) = \frac{8 \pi f^2}{M_\pi [1-(M_\pi^2/4 m_p^2)]} + 4 \pi M_\pi J^- 
= 4 \pi (1+x) a^-_{0+} \, ,
\end{eqnarray}
where
\begin{eqnarray}
J^- = \frac{1}{2 \pi^2} \int_0^{\infty} \frac{\sigma^-(k)}{\omega} \; dk
\end{eqnarray}
and $x=M_\pi/m_p$.
The pion-nucleon coupling constant can now be extracted and the result is
\begin{eqnarray}
f^2 &=& \frac{1}{2} [1-(\frac{x}{2})^2]
[(1+x) M_\pi a^-_{0+} - M_\pi^2 J^-] \nonumber \\
&=& 0.5712 (M_\pi \, a^-_{0+}) -0.02488 (J^-/{\rm mb}).
\end{eqnarray}
The isovector s-wave scattering length, $a^-_{0+}$, is accessible
through experiment \cite{19}. For the integral $J^-$ several evaluations
are displayed in Table 1. As can be seen from Fig. 3 there is potential
sensitivity to details of the electromagnetic corrections especially
around the $\Delta$-resonance region near 0.3 GeV/c as well as to the treatment
of the $\Delta^{++}$, $\Delta^0$ splitting.
\begin{table}[h]
\begin{center}
\caption{Values for the $J^-$ integral.}
\begin{tabular}{l|l}
Ref. & $J^-$ (mb) \\ \hline
KH ('83) \cite{2}& -1.058 \\
Koch ('85)\cite{15} & -1.077 $\pm$ 0.047 \\
VPI ('92) \cite{16}& -1.072 \\
Gibbs ('98) \cite{17}& -1.051  \\
ELT ('99) \cite{18}& -1.083 $\pm$ 0.025 \\ \hline
\end{tabular}
\end{center}
\end{table}
Making use of the isospin symmetry gives for the scattering length
$a^-_{0+}=0.0962 \pm 0.0071 \, M^{-1}_\pi$ \cite{19} and taking Koch's value
for $J^-$ gives an estimate for the lower limit of the coupling constant $f^2$
with the result 0.0765. With more conservative errors for $J^-$ the figure
0.0762 is obtained. With the remaining uncertainties in the treatment of various
corrections this limit is not in real conflict with the results from other
analyses.

\begin{figure}[t]
\begin{center}
\mbox{\epsfysize=8cm \epsfbox{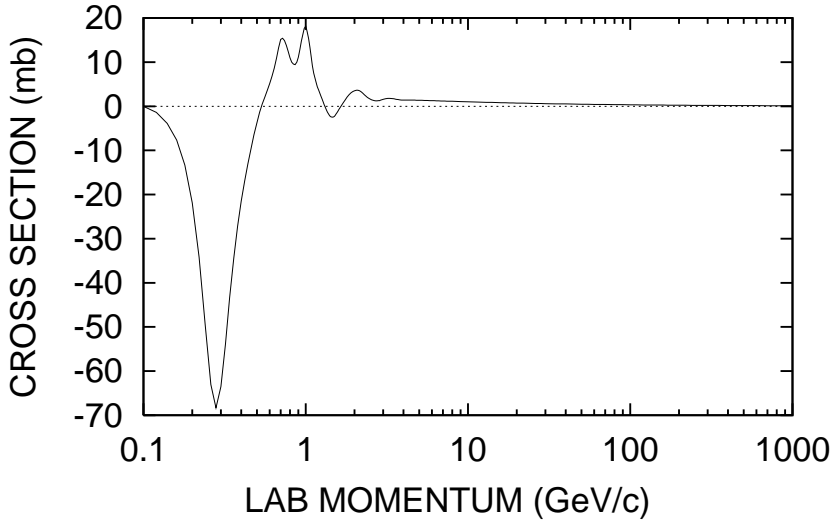}
     }
\caption{The isovector combination, $\sigma^- = \frac{1}{2}(\sigma_{\pi^-p}-\sigma_{\pi^+p})$,
of the $\pi^-p$ and $\pi^+p$ total cross sections \cite{2}. Experimental data
extend up to 350 GeV/c.}
\label{fig3} \end{center} \end{figure}

\section*{CONCLUDING REMARKS}

The precision of the pion-nucleon experiments has now reached the level
where a more careful treatment of the corrections, in particular of
electromagnetic origin or due to the $u$- and $d$-quark mass difference,
is necessary. These theoretical challenges have not yet been met, eventhough
the theoretical tool, chiral perturbation theory, has now the capability
to answer these questions. Work along these lines is in progress.
In the lattice \cite{20} and the QCD sum rule \cite{21} frontiers the present accuracy
for $g$ is 20-30 \% and it will take a while before this improves significantly.

Table 2 summarizes some recent values for $f^2$ displayed in Fig. 2.
\begin{table}[h]
\begin{center}
\caption{Values for the pion-nucleon coupling constant $f^2$ from recent determinations.}
\begin{tabular}{l|l|l}
Ref. & $f^2$ & Method \\ \hline
KH ('80) \cite{7}& 0.079 $\pm$ 0.001 & $\pi$N fixed-$t$ \\
BM ('95) \cite{22} & 0.0757 $\pm$ 0.0022 & NN data \\
Gibbs ('98) \cite{17} & 0.0756 $\pm$ 0.0007 & GMO \\
Machner ('98) \cite{23} & 0.0760 $\pm$ 0.0011 & symmetries \\
Matsinos ('98) \cite{24} & 0.0766 $\pm$ 0.0011 & model fit \\
Nijmegen ('99) \cite{25} & 0.0756 $\pm$ 0.0004 & $pp$ PWA \\
ELT ('99) \cite{18}& 0.0786 $\pm$ 0.0008 & GMO + $\pi^-d$ \\ 
VPI ('99) \cite{10,26} & 0.0760 $\pm$ 0.0004 & $\pi$N fixed-$t$ \\\hline
\end{tabular}
\end{center}
\end{table}
The table demonstrates the current trend, the favoured value for $f^2$ is slightly
smaller than the standard one of Koch and Pietarinen \cite{7}. However, there
remains still quite a number of problems which need attention. 

The Goldberger-Treiman discrepancy
\begin{eqnarray}
\Delta_{\pi N} = 1-\frac{m_p \, g_A}{F_\pi \, g},
\end{eqnarray}
where $F_\pi$ and $g_A$ are the pion and neutron decay constants respectively,
would be reduced from 4 \% to 2 \%, if $f^2$ changes from 0.079 to 0.076. 
In a recent SU(3) analysis \cite{27} preference for a smaller Goldberger-Treiman
discrepancy was found.

In the analysis of $np$ scattering data at backward directions somewhat
higher value for the coupling constant has been obtained \cite{28}, 
$f^2$ = 0.0803 $\pm$ 0.0014. Discussion on the problems in this field 
continues \cite{29,30}. The spin transfer coefficients in $pp$ scattering
are also of interest, the preliminary indications are towards slightly
smaller value for the coupling \cite{31}. Machleidt has recently 
discussed \cite{32} some additional problems with the deuteron properties 
and low-energy NN analyzing powers which indicate that no coherent picture
is yet emerging.

\section*{ACKNOWLEDGEMENTS}

I thank A.M. Green for useful remarks on the manuscript. Partial support
from the EU-TMR programme, contract CT98-0169, is gratefully acknowledged.

\bibliographystyle{unsrt}

\begin{thebibliography}{99}
\bibitem{32} R.~Machleidt, ``How sensitive are various NN observables to changes
in the $\pi$NN coupling constant?,'' nucl-th/9909036.
\end{thebibliography}

\end{document}